\documentclass[twoside]{dis07}
\usepackage[latin1]{inputenc}
\usepackage[dvips]{graphicx,epsfig,color}
\usepackage{wrapfig,rotating}
\usepackage{amssymb,amsmath,array}
\usepackage{caption}

\begin{document}
\title{Exclusive $\rho^0$ electroproduction}

\author{Aharon Levy$^{1,2}$ on behalf of the ZEUS collaboration
%
%
\vspace{.3cm}\\
%
1- Raymond and Beverly Sackler Faculty of Exact Sciences, School of
Physics and Astronomy, \\ Tel Aviv University, Tel Aviv, Israel
%
\vspace{.1cm}\\
2- DESY, Hamburg, Germany
}

\maketitle

\begin{abstract}
{Exclusive $\rho^0$ electroproduction at HERA has been studied with
the ZEUS detector, using 120 pb$^{-1}$ integrated luminosity, in the
kinematic range of photon virtuality of $2 < Q^2 < 160$ GeV$^2$, and
$\gamma^* p$ center-of-mass energy of $32 < W < 180$ GeV.  The results
include the $Q^2$ and $W$ dependence of the $\gamma^* p
\to \rho^0 p$ cross section and the distribution of the
squared-four-momentum transfer to the proton, $t$.  Also included is
the ratio of longitudinal to transverse $\gamma^* p$ cross section as
a function of $Q^2$, $W$ and $t$. Finally, the effective Pomeron
trajectory was extracted.  The results are compared to various
theoretical predictions, none of which are able to reproduce all the
features of the data.  }
\end{abstract}


Exclusive electroproduction of light vector mesons is a particularly
good process for studying the transition from the soft to the hard
regime, the former being well described within the Regge phenomenology
while the latter - by perturbative QCD. Among the most striking
expectations in this transition is the change of the logarithmic
derivative $\delta$ of the cross section $\sigma$ with respect to the
$\gamma^* p$ center-of-mass energy $W$, from a value of about 0.2 in
the soft regime to 0.8 in the hard one, and the decrease of the
exponential slope $b$ of the differential cross section with respect
to the squared-four-momentum transfer $t$, from a value of about 10
GeV$^{-2}$ to an asymptotic value of about 5 GeV$^{-2}$ when the
virtuality $Q^2$ of the photon increases.

In this talk, the latest results of a high statistic measurement of
the reaction $\gamma^* p \to \rho^0 p$ studied with the ZEUS detector
are presented. A detailed presentation can be found
in~\cite{url}. Here we present the main results.


\begin{wrapfigure}{r}{0.5\columnwidth}
\vspace{-1cm}
\centerline{\includegraphics[width=0.45\columnwidth]{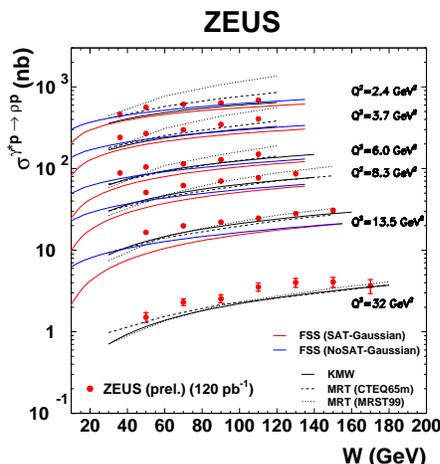}}
\vspace{-0.3cm}
\caption{\it $W$ dependence of $\sigma$ for different values of $Q^2$. The lines are 
the predictions of some models (see text).}
\label{fig:wdep-th}
\end{wrapfigure}
\begin{figure}
\begin{minipage}[h]{0.5\columnwidth}
\hspace{-5mm}
{\includegraphics[width=1.05\columnwidth]{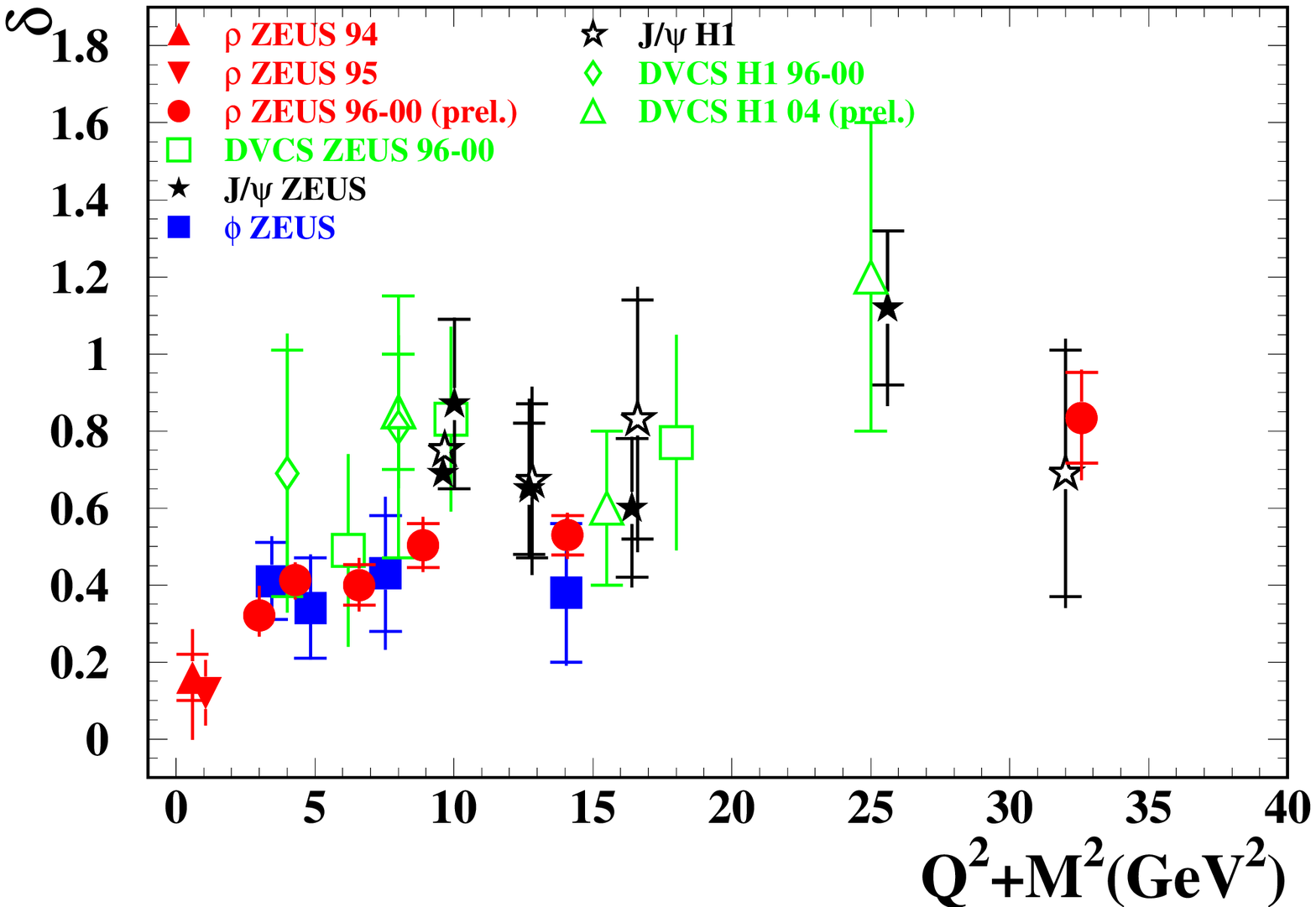}}
\caption{\it $\delta$ as a function of $Q^2+M^2$.}
\label{fig:del07}
\end{minipage}
\hspace{5mm}
\begin{minipage}[h]{0.5\columnwidth}
\hspace{-5mm}
{\includegraphics[width=1.05\columnwidth]{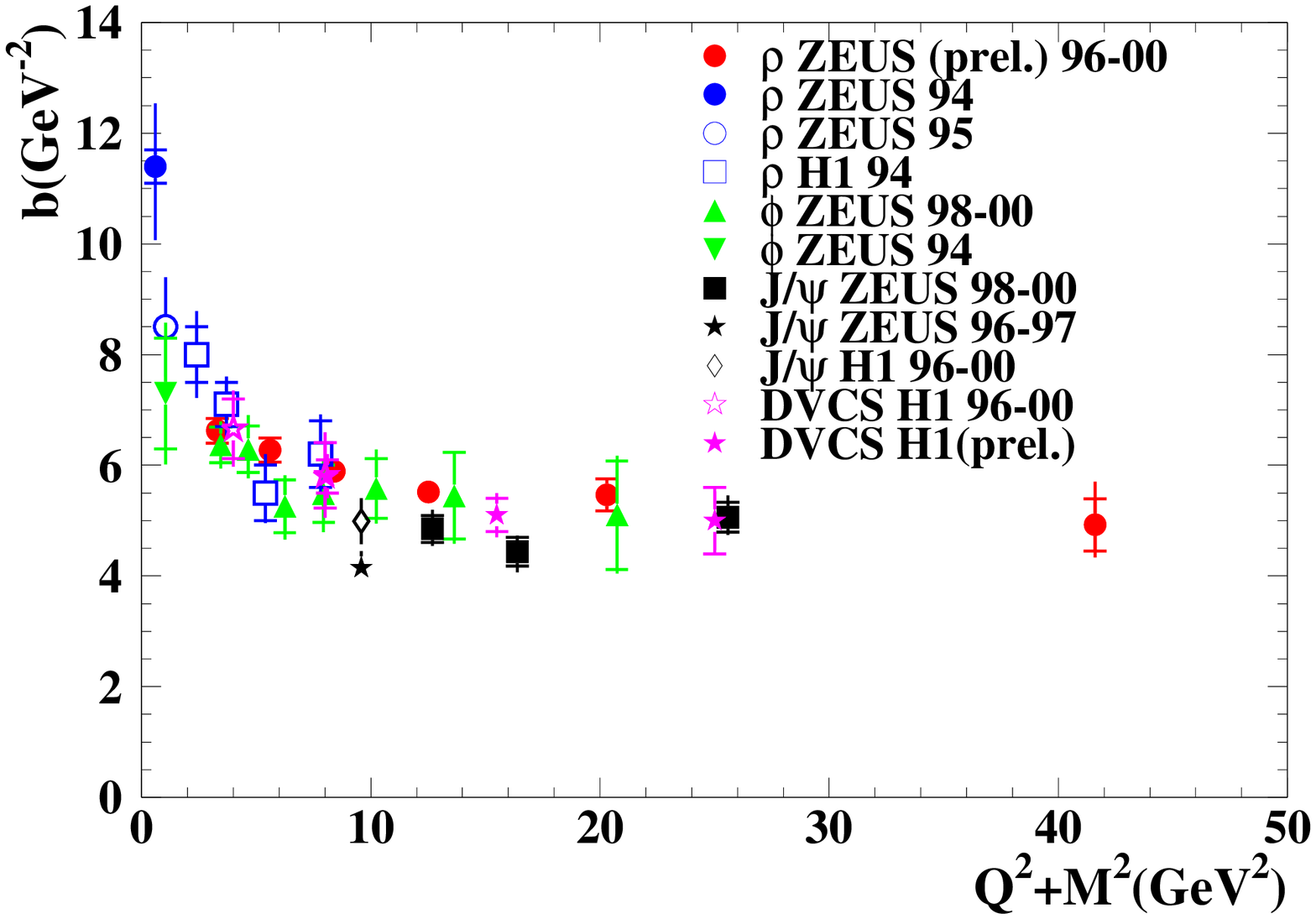}}
\caption{\it $b$ as a function of $Q^2+M^2$.}
\label{fig:b07}
\end{minipage}
\end{figure}

The cross section $\sigma (\gamma^* p \to \rho^0 p)$ is presented in
Fig.~\ref{fig:wdep-th} as a function of $W$, for different values of
$Q^2$. The cross section rises with $W$ in all $Q^2$ bins.  In order
to quantify this rise, the logarithmic derivative $\delta$ of $\sigma$
with respect to $W$ is obtained by fitting the data to the expression
$\sigma \sim W^\delta$ in each of the $Q^2$ intervals. The resulting
values of $\delta$ are shown in Fig~\ref{fig:del07}.  Also included in
this figure are values of $\delta$ from lower $Q^2$ measurements for
the $\rho^0$ as well as those for $\phi$, $J/\psi$ and $\gamma$
(Deeply Virtual Compton Scattering (DVCS)). In this case the results
are plotted as function of $Q^2+M^2$, where $M$ is the mass of the
vector meson.  One sees a universal behaviour of the different
processes, showing an increase of $\delta$ as the scale becomes
larger, in agreement with the expectations mentioned above.  The value
at low scale is the one expected from the soft Pomeron
intercept~\cite{dl}, while the one at large scale is in accordance
with twice the logarithmic derivative of the gluon density with
respect to $W$.

The differential cross section, d$\sigma$/d$t$, has been parameterised
by an exponential function $e^{-b|t|}$ fitted to the data. The
resulting values of $b$ as a function of the scale $Q^2+M^2$ are
plotted in Fig.~\ref{fig:b07} together with those from other
processes. As expected, $b$ decreases to a universal value of about 5
GeV$^{-2}$ as the scale increases. This value measures the radius of
the gluon density in the proton and corresponds to a value of $\sim$
0.6 fm, smaller than the value of the charge density of the proton
($\sim$ 0.8 fm), indicating that the gluons are well-contained within
the charge-radius of the proton.


\begin{wrapfigure}{r}{0.5\columnwidth}
\vspace{-1.3cm}
\hspace{0.4cm}
\centerline{\includegraphics[width=0.45\columnwidth]{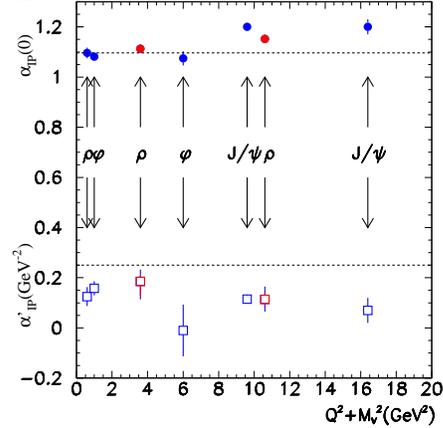}}
\vspace{-0.6cm}
\caption{\it A compilation of $\alpha_{IP}(0)$ and $\alpha_{IP}^\prime$ for $\rho$, $\phi$ and 
$J/\psi$, as a function of $Q^2+M^2$.}
\label{fig:ap-apr-pom}
\end{wrapfigure}
One can study the $W$ dependence of d$\sigma$/d$t$ for fixed $t$
values and extract the effective Pomeron trajectory
$\alpha_{IP}(t)$. This was done for two $Q^2$ intervals and the
trajectory was fitted to a linear form to obtain the intercept
$\alpha_{IP}(0)$ and the slope $\alpha_{IP}^\prime$, the values of
which are tabulated in Table~\ref{tab:alpha}.  A compilation of the
effective Pomeron intercept and slope from this study together with
that from other vector mesons is presented in
Fig.~\ref{fig:ap-apr-pom}. As in the other compilations, the values
are plotted as a function of $Q^2+M^2$.  The value of $\alpha_{IP}(0)$
increases slightly with $Q^2$ while the value of $\alpha_{IP}^\prime$
shows a small decrease with $Q^2$.

\begin{table}[hbt]
\begin{center}
\begin{tabular}{|c|c|c|c|}
\hline
$Q^2$ (GeV$^2$) & $<Q^2>$ (GeV$^2$) &  $\alpha_{IP}(0)$ &  $\alpha_{IP}^\prime$(GeV$^{-2}$ \\
\hline
$2-5$  & $3$ & $1.113\pm 0.010^{+ 0.009}_{- 0.012}$ &  $0.185\pm 0.042^{+ 0.022}_{- 0.057}$  \\
$5-50$ & $10$ & $1.152\pm 0.011^{+ 0.006}_{- 0.006}$ &  $0.114\pm 0.043^{+ 0.026}_{- 0.024}$  \\
\hline
\end{tabular}
\caption{\it
The values of the Pomeron trajectory intercept $\alpha_{IP}(0)$ and
slope $\alpha_{IP}^\prime$, for different $Q^2$ intervals.  }
\label{tab:alpha}
\end{center}
\end{table}
\hspace{-5mm}

\begin{figure}[h]
\begin{minipage}{0.5\columnwidth}
\hspace{-0.5cm}
\centerline{\includegraphics[width=\columnwidth]{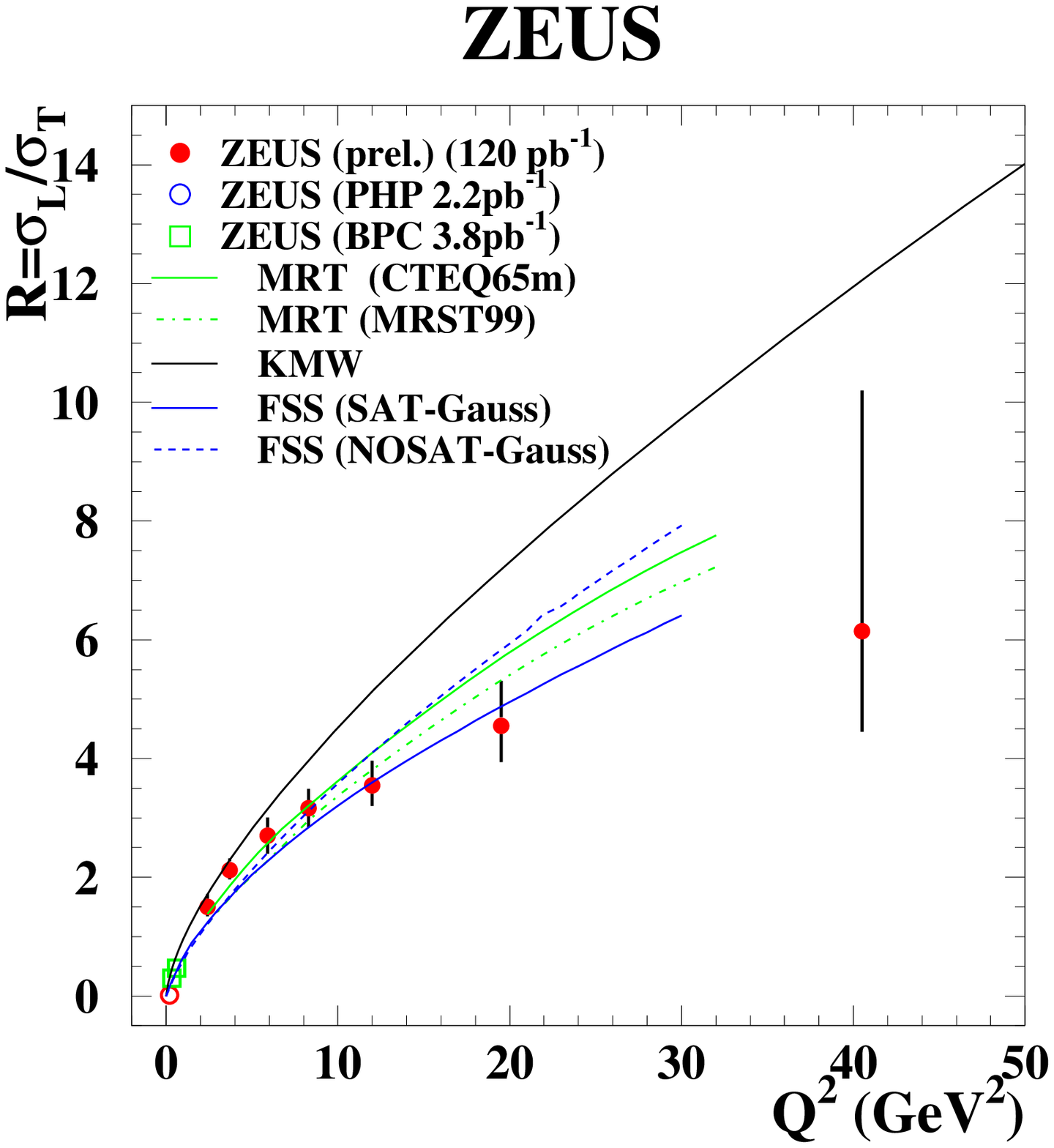}}
\caption{\it $R$ as a function of $Q^2$ at $W$=90 GeV. The lines are the 
prediction of models referred to in the text.}
\label{fig:r-q2}
\end{minipage}
\hspace{2mm}
\begin{minipage}[h]{0.5\columnwidth}
\hspace{-0.5cm}
\centerline{\includegraphics[width=\columnwidth]{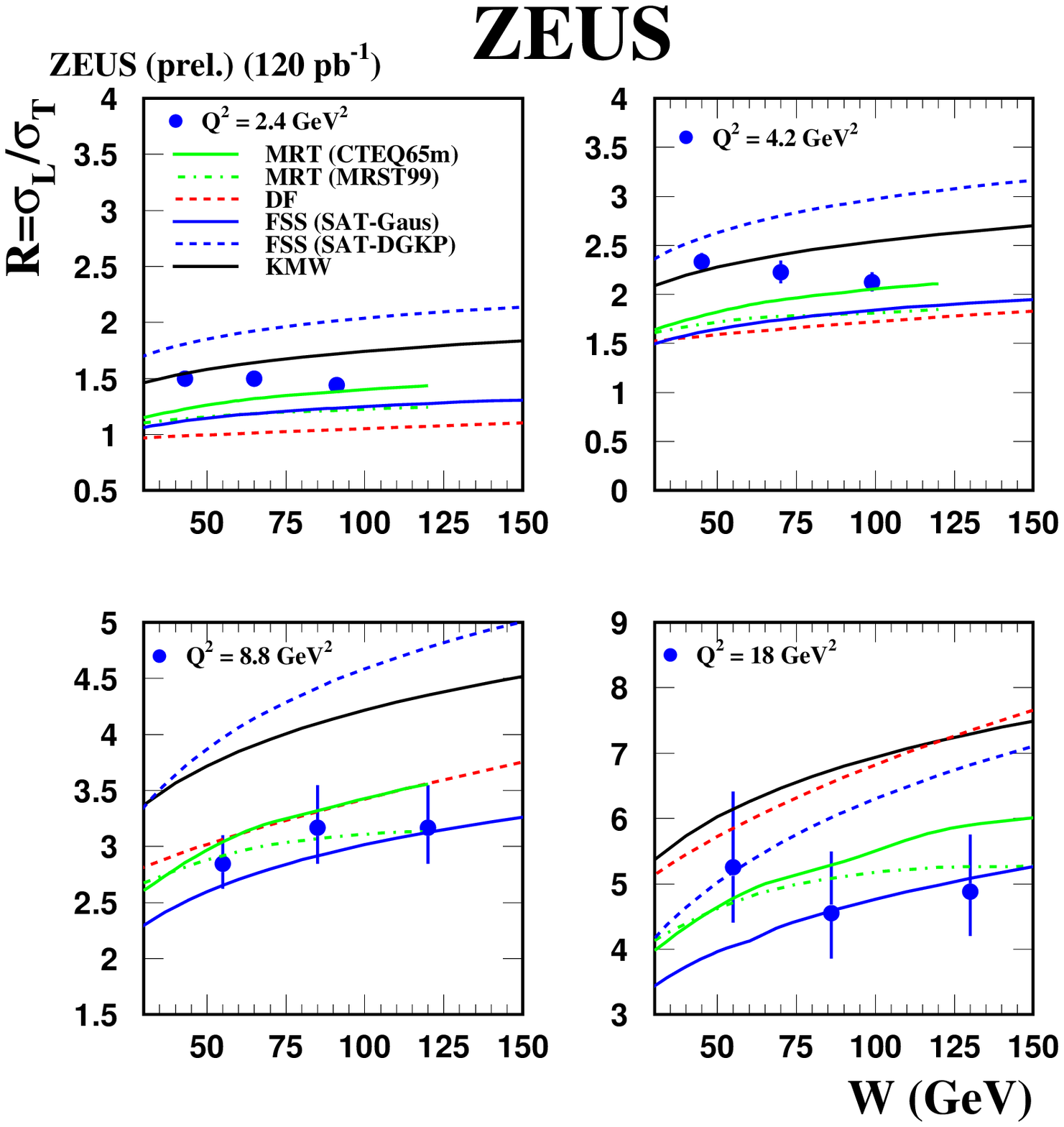}}
\caption{\it The ratio $R$ as a function of $W$ for different $Q^2$ intervals. 
The lines are the prediction of models
referred to in the text.}
\label{fig:r-w-th}
\end{minipage}
\end{figure}

The helicity analysis of the decay-matrix elements of the $\rho^0$ was
used to extract the ratio $R$ of longitudinal to transverse $\gamma^*
p$ cross section, as a function of $Q^2$, $W$ and $t$. While $R$ is an
increasing function of $Q^2$, as shown in Fig.~\ref{fig:r-q2}, it is
independent of $W$ in all $Q^2$ intervals
(Fig.~\ref{fig:r-w-th}). This unexpected behaviour indicates that the
large configurations in the wave function of the transverse $\gamma^*$
seem to be suppressed. This result is supported by the independence of
$R$ on $t$ (not shown), indicating that both polarisations of the
photon fluctuate into similar size $q\bar{q}$ pairs.

The results of this study were compared to those of the H1
collaboration~\cite{h1-disrho} and both measurements are in good
agreement.

The results were also compared to several theoretical predictions. The
predictions are a combination of perturbative and non-perturbative QCD
calculations. All models use the dipole picture to describe the
reaction $\gamma^* p \to \rho^0 p$. The ingredients necessary for the
calculation are the virtual photon and the $\rho^0$ wave function and
the gluon densities. Some models put their emphasis on the VM wave
function~\cite{fks,fss,kmw,df} while that of~\cite{mrt} studies the
dependence on the gluon densities in the proton. Detailed comparison
can be seen in~\cite{url}. Some examples are shown in
Fig.~\ref{fig:wdep-th}, where the cross section values are plotted as
a function of $W$ for different $Q^2$ vales, in Fig.~\ref{fig:r-q2},
where the ratio $R$ is shown as a function of $Q^2$ and in
Fig.~\ref{fig:r-w-th}, where $R$ is plotted as a function of $W$, for
different $Q^2$ intervals. As can be seen, none of the calculations
can describe the data.  The high precision of the present measurements
can be most valuable to tune the different models and thus contribute
to a better understanding of the $\rho^0$ wave function and of the
gluon density in the proton.

\section*{Acknowledgments}

It is a pleasure to acknowledge and thank Allen Caldwell and his local
organizing committee for a superb organization of this workshop. This
work was partly supported by the Israel Science Foundation (ISF).


\begin{footnotesize}


\end{footnotesize}


\end{document}